# The Status Quo of Architecture and Its Impact on Urban Management: Christopher Alexander's Insights

Bin Jiang
LivableCityLAB, Thrust of Urban Governance and Design,
The Hong Kong University of Science and Technology (Guangzhou), China
Email: binjiang@hkust-gz.edu.cn
*(Draft: August 2022, Revision: July, August, and September 2024)*

**Abstract**
Christopher Alexander offers a critical perspective on the modernist approach to architecture, which he argues has prioritized innovation, abstraction, and mechanistic efficiency at the expense of human-centered and organic values. This shift has led to the proliferation of buildings that, while visually striking, often feel cold, impersonal, and disconnected from the deeper needs of the people who inhabit them. Alexander advocates for a return to timeless architectural principles such as harmony, balance, and a deep connection to the natural and cultural context. He introduces the concept of living structure, which emphasizes creating spaces that resonate with the intrinsic order found in nature and human life, fostering environments that are not only functional and beautiful but also profoundly life-affirming. By challenging the dominance of "iconic" but alienating designs, Alexander calls for a holistic, human-centered approach to architecture—one that prioritizes the well-being of individuals and communities, creating spaces that nurture a sense of place, belonging, and harmony with the world around them.

## 1. Introduction

The architectural landscape of the 20th and 21st centuries has been marked by a profound and pervasive crisis, one that Christopher Alexander, a seminal figure in architectural theory (Mehaffy 2007a, 2007b, Salingaros 2020), has vocally critiqued. This crisis strikes at the core of architectural practice and education. In the aftermath of modernism's rise, architecture has increasingly been driven by principles that prioritize novelty, abstraction, and mechanistic efficiency (Jencks 2002) over humanistic and organic values. The shift toward sleek, often sterile forms has led to buildings that, while visually innovative, fail to engage with the deeper aspects of human experience and environmental harmony. This architectural crisis mirrors broader challenges in urban management, where cities increasingly prioritize innovation and efficiency at the expense of human-centered values. This paper situates Alexander's insights within the urban context, proposing that the principles he advocates for—such as harmony, balance, and living structure—are crucial for shaping cities that are both functional and socially sustainable.

Christopher Alexander's critique of contemporary architecture (Alexander 2002–2005) is both scathing and insightful. He contends that the modernist movement, with its emphasis on breaking from tradition and embracing industrialization and technology as ends in themselves, has led to a disconnection between architecture and its fundamental purpose: To create spaces that nurture, sustain, and enhance human life. The modernist agenda, with its focus on form over substance and on the aestheticization of the machine, has produced an architectural landscape dominated by buildings that may be visually striking but are often devoid of the qualities that make spaces feel alive and meaningful.

Alexander's insights challenge the very foundations upon which modernist practices are built. He questions the validity of an architectural approach that views buildings as mere objects to be admired for their visual "impact", rather than as living entities that must interact harmoniously with their environment and the people who inhabit them. According to Alexander, the crisis in architecture is rooted in a fundamental misunderstanding of what it means to create spaces that are truly beautiful and meaningful. Beauty, in Alexander's view, is not a superficial attribute but a manifestation of deeper, more complex relationships between a space and its users, between form and function, and between the built environment and the natural world.

His diagnosis of the ailments afflicting contemporary architecture is both a critique and a call to action. Alexander advocates for a profound rethinking of discipline, one that moves away from the mechanistic, dehumanized approaches that have dominated the field for the past century (Le Corbusier 1923, Millais 2017). Instead, he calls for a return to the principles that have historically governed the creation of architecture—principles that emphasize harmony, balance, and a deep connection to the natural and cultural context (Alexander 1979). These principles are not merely nostalgic but are rooted in a timeless understanding of what makes spaces meaningful and alive.

In advocating for a return to these principles, Alexander is not merely calling for a revival of past styles or forms (Alexander 2002). Instead, he seeks a reinvigoration of the underlying values that have always made architecture a vital and life-affirming discipline. His vision for architecture is deeply rooted in the recognition of the intrinsic order and beauty inherent in the natural world, emphasizing that the built environment must resonate with these qualities to truly serve humanity. This paper reviews the current architectural crisis and the rise of the grassroots movement toward new traditional architecture. We introduce the concept of living structure and critically examine modernist architecture, with a focus on starchitects. Finally, we present a new vision of architecture rooted in living structure, framed within an organic view of space and the world.

The remainder of this paper is organized as follows. Section 2 addresses the architectural crisis, examining how modernist architecture has strayed from holistic design principles. Section 3 explores the grassroots movement calling for a new traditional architecture, advocating for a return to enduring design values. Section 4 introduces Alexander's concept of living structure, proposing a more integrated and life-affirming approach to architecture. Section 5 critiques the role of "starchitects" and the focus on novelty over substance, while Section 6 presents a holistic vision for the future of architecture. Finally, Section 7 draws on Alexander's insights and concludes with the implications.

**2. The Architectural Crisis**
The crisis in contemporary architecture, as observed by Alexander (2002–2005), originates from a profound departure from the holistic and integrative principles that historically guided the discipline and architectural practices. In earlier architectural traditions, design was an organic process where every element of a building was conceived in relation to the whole, ensuring that the final structure was not only functional but also imbued with a sense of harmony and life (Venturi 1966). This approach recognized the importance of creating spaces that were in tune with both their physical environment and the psychological needs of their inhabitants.

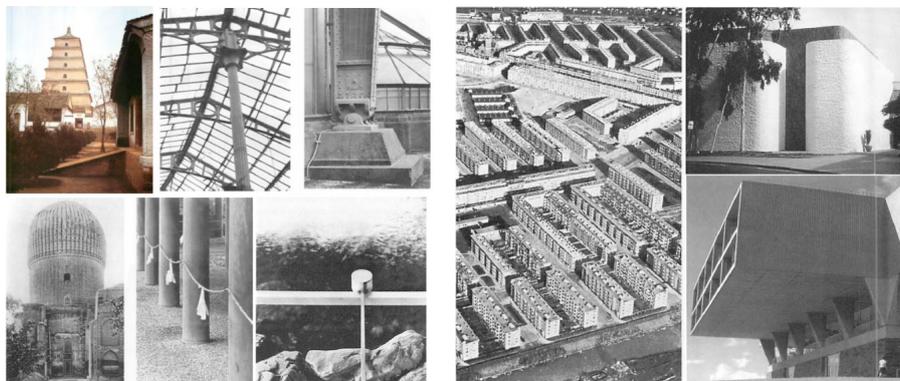

Figure 1: Two types of architecture imply two kinds of order: Organic and mechanical (Alexander 2002–2005)

The advent of modernist architecture in the 20th century, however, marked a significant shift in this paradigm. Modernism, driven by rapid technological advancements and the desire to break away from historical precedents, embraced a mechanistic worldview (Curl 2018). This perspective treated buildings as machines, reducing architecture to a series of functional components that could be assembled and manipulated with little regard for their relationship to the broader environment or to human experience (c.f. Figure 1 for two types of architecture). The focus shifted towards efficiency, standardization, and a minimalist aesthetic that favored stark forms and abstract geometries over the warmth and richness of traditional designs

This mechanistic approach led to a prioritization of form over function, where the visual "impact" of a building often took precedence over its utility and livability. Architects became more concerned with creating "iconic" structures that could stand out in a crowded urban landscape, often at the expense of the human experience. The result was a proliferation of buildings that, while visually striking, often lacked the deeper qualities that contribute to a sense of place and belonging (Lewicka 2011). These structures, characterized by their sleek lines and bold forms, frequently appeared cold, impersonal, and disconnected from the needs of the people who used them.

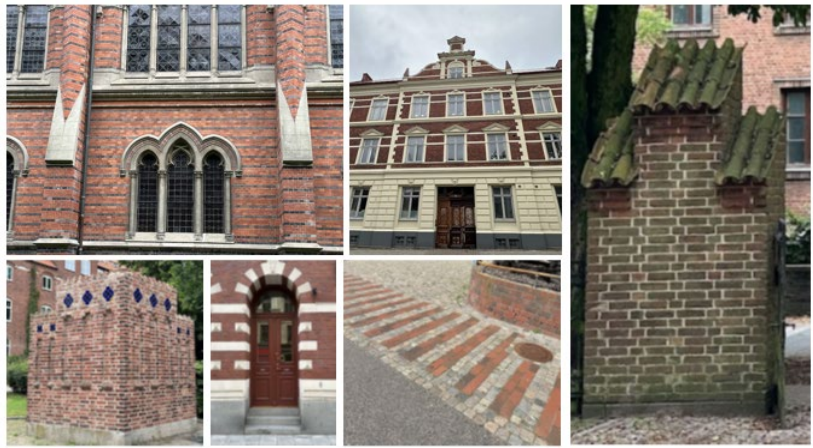

Figure 2: Examples of brickwork and organic architectural structures from the past

Alexander describes this phenomenon as a form of "ugliness" that has come to dominate our cities and towns of the 20th and 21st centuries. This ugliness is not merely a matter of aesthetic preference; it reflects a deeper malaise in the way we conceive and create our built environment. Instead of spaces that nurture and support human life (Alexander 1979), we have constructed environments that are often alienating and devoid of the qualities that make places feel alive and welcoming. The root of this ugliness lies in the detachment of architectural practice from the principles of harmony, order, and beauty that once guided the creation of human-centered spaces.

Furthermore, this architectural crisis is exacerbated by the growing influence of commercial interests in the design process. In a world where economic considerations often dictate design choices, the creation of spaces has become increasingly driven by market forces and superficial trends. Developers and architects, in their quest for profitability and recognition, may opt for designs that are flashy and attention-grabbing but ultimately fail to meet the deeper needs of the people who inhabit them. The result is a built environment that prioritizes short-term visual appeal and commercial success over long-term sustainability and human well-being.

This crisis is evident in the way many modern buildings and urban development feel disconnected from their surroundings and the people who use them. The homogenization of design, with its reliance on standardized materials and construction techniques, has led to the creation of spaces that lack character and fail to resonate with the unique identities of the communities they serve. In contrast to the richly textured and contextually sensitive architecture of the past (Figure 2), much of modern architecture seems sterile and out of place, contributing to a sense of disorientation and fragmentation in our cities.

Alexander's critique of modernist architecture is a call to return to a more human-centered approach to design, one that recognizes the importance of creating spaces that are not only functional but also deeply connected to the lives and experiences of the people who use them. He advocates for a reengagement with the principles of harmony, order, and beauty, suggesting that only by restoring these values to their rightful place at the heart of architectural practice can we begin to address the crisis that has engulfed the discipline.

This crisis in architecture is reflected in the design of modern cities. The prioritization of sleek, mechanistic structures over organic and human-centered design has contributed to urban environments that lack vitality and fail to foster a sense of place. As cities grow, the need for a more holistic, human-centered approach to urban design and management becomes increasingly urgent.

### 3. The Movement Towards a New Traditional Architecture

In response to the perceived shortcomings of modernism, a burgeoning movement known as "new traditional architecture" has emerged, championed by figures like Michael Diamant (2008). This movement represents a conscious rejection of the dominant architectural trends of the 20th and 21st centuries, which often prioritized innovation, abstraction, and stylistic experimentation at the expense of timeless design principles (Alexander 1979). Proponents of new traditional architecture argue that the aesthetic and functional dissonance found in much of modernist architecture has contributed to what they describe as the "uglification" of urban environments. This term encapsulates the proliferation of buildings that, while often celebrated for their novelty, fail to resonate with the deeper human need for beauty, order, and harmony.

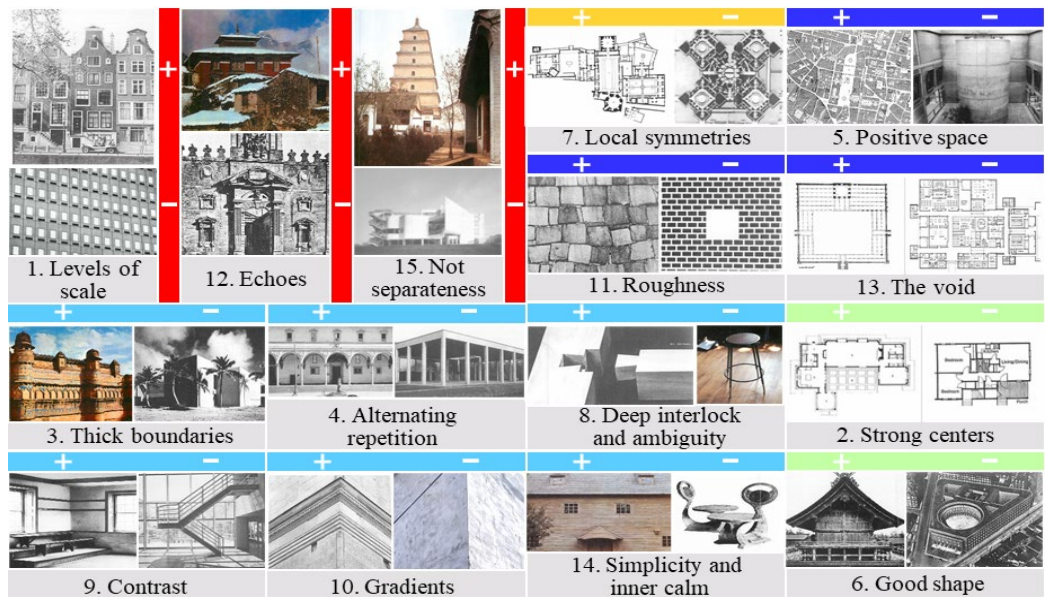

Figure 3: (Color online) Visualization of the fifteen geometrical and transformation properties
(Note: This diagram illustrates the fifteen properties of living structure, each shown with positive (+) and negative (-) examples, mainly from Alexander (2002–2005). The properties are divided into five groups, each color-coded. The red group represents global properties spanning multiple scales, while the two blue groups highlight local properties. 'Local symmetries' is uniquely emphasized, and 'strong centers' and 'good shape' are included as optional, practical properties, though not part of the core fifteen.)

At its core, new traditional architecture is a call to return to the classical principles that have guided architectural practice for millennia. These principles, rooted in the traditions of ancient Greece and Rome and refined through the Renaissance and other periods of architectural flourishing, emphasize the importance of proportion, scale, and harmony in design. Similarly, traditional architectures from diverse cultural heritages, such as Chinese, Japanese, Korean, African, Spanish, and other indigenous and regional styles developed over centuries, embody their own expressions of these principles, often incorporating elements of symmetry, spatial balance, and natural harmony that reflect the values, beliefs, and environmental contexts of their respective societies. Unlike modernist architecture, which often seeks to break from the past and create entirely new forms, new traditional architecture embraces historical precedents as a source of enduring wisdom. The movement asserts that the classical approach holds intrinsic value, fostering a universally recognizable sense of balance and coherence that is deeply comforting. This is exemplified by the 15 geometrical and transformational properties (Figure 3), which are believed to be universal across all architectural traditions and cultural contexts.

One of the primary criticisms of modernism by proponents of new traditional architecture is its tendency to create spaces that feel alienating and disjointed. The stark, minimalist forms and monolithic structures often come across as cold, impersonal, and disconnected from both the human scale and the natural environment. In contrast, new traditional architecture aims to design buildings that are not only functional but also deeply connected to human emotions and experiences. This connection arises from the incorporation of living structures, which evoke a strong sense of place attachment (Lewicka 2011). By employing architectural principles such as levels of scale, local symmetries, thick boundaries, and alternating repetition, new traditional architecture fosters environments that feel harmonious and emotionally resonant.

Furthermore, the movement emphasizes the importance of creating spaces that are "alive"—that is, spaces that resonate with the human spirit and contribute to the well-being of their occupants. This concept of "aliveness" is closely related to the concept of "living structure" (Alexander 2002–2005), where the built environment is designed in a way that supports human life in its fullest sense, encouraging interaction, reflection, and a sense of belonging. New traditional architecture, therefore, is not merely about replicating the past but about drawing on timeless principles (Alexander 1979) to create environments that are vibrant and responsive to the needs of contemporary society.

The resurgence of traditional architectural practices can also be seen as a form of resistance against the homogenization of global architecture. In an era where international styles often dominate, leading to cities around the world looking increasingly alike, new traditional architecture advocates for a more localized, culturally informed approach. By re-engaging with regional architectural traditions, this movement seeks to restore a sense of place and identity to the built environment (Lewicka 2011), ensuring that architecture reflects the values, history, and character of the communities it serves.

In essence, the movement towards new traditional architecture is both a critique of the modernist paradigm and a proactive effort to reclaim the lost art of building. It invites architects to look beyond the transient allure of novelty and to rediscover the enduring qualities of beauty, proportion, and harmony that have defined great architecture throughout history. Through this approach, new traditional architecture aims to create spaces that are not only aesthetically pleasing but also deeply resonant with the human condition, fostering environments that nurture both the body and the soul.

**4. Living Structure: A New Perspective on Architecture**
The concept of living structure introduces a transformative way of thinking about architecture and urban management, one that moves beyond the mechanistic and reductionist views that have long dominated the discipline. This concept suggests that all matter—whether organic or inorganic—possesses a degree of life, not merely in a biological sense but as an intrinsic quality of its structure and arrangement (Alexander 2002–2005, Gabriel and Quillien 2019, Jiang 2019). This idea challenges the conventional understanding of architecture as simply the assembly of materials to meet functional requirements (Jiang and Huang 2021). Instead, it invites us to consider the deeper, often intangible qualities that make a space feel alive, resonant, and meaningful. In urban environments, living structure encourages the creation of cities that grow organically and remain adaptable over time. It emphasizes the importance of interconnectedness, both within the built environment and between the city and its natural surroundings. By applying living structure principles, urban managers and planners can create spaces that support social interaction, foster a sense of community, and enhance the overall well-being of their residents.

The notion of living structure is a direct response to the mechanistic worldview that has heavily influenced modern architecture. In this view, buildings are often conceived as machines, composed of discrete parts that operate independently to serve specific functions. While this approach has led to innovations in efficiency and the optimization of space, it has also resulted in designs that can feel cold, alienating, and disconnected from their environment. Alexander argues that such structures, while they may meet functional and aesthetic criteria on the surface, often lack the deeper sense of life that makes a space truly engaging and supportive of human well-being.

The living structure concept posits that spaces and buildings can be designed to resonate with their surroundings, creating a sense of harmony, coherence, and vitality. This resonance is not just a matter of visual appeal; it is a quality that can be both objectively measured and subjectively felt. For instance, a living structure might incorporate patterns, proportions, and relationships that echo those found in nature, fostering a deep connection between the built environment and the natural world (c.f. Figure 4 for two examples: one with a more dynamic, living structure, and the other with a less vibrant, more static design). This approach requires us to consider every element of a building, from the smallest detail to the overall form, as part of an interconnected whole that contributes to the life of the structure.

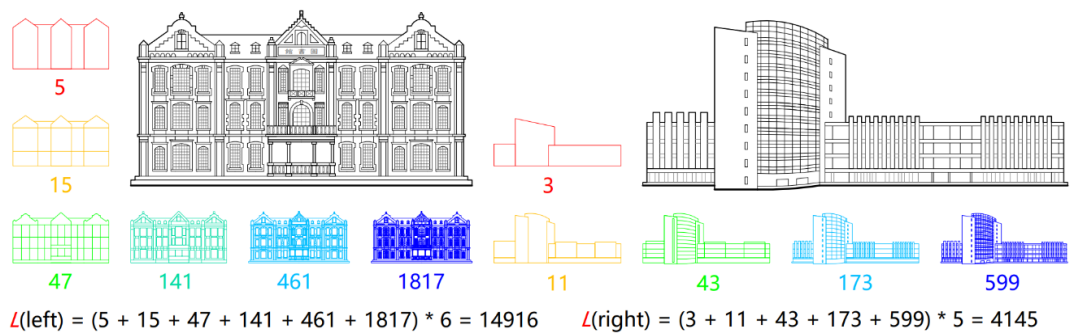

Figure 4: (Color online) Shanghai Jiao Tong University library buildings and their substructures (Note: The degree of living structure is calculated using the formula $L = S \times H$ (Jiang and de Rijke 2023). The older library on the left exhibits a higher degree of living structure than the newer one on the right.)

One of the key principles Alexander emphasizes is the importance of 'centers' or substructures as referred to by this author (c.f. Figure 4 for substructures) within a design. These are focal points that naturally attract attention and organize the space around them. In a living structure, centers are not isolated elements but are part of a network of relationships that give the building its sense of coherence and vitality. The degree of life in a structure can be assessed by how well these centers are integrated and how they contribute to the overall harmony of space.

The implications of living structure extend far beyond individual buildings. Alexander's ideas challenge us to think about the broader context in which their designs and plans exist. A building that embodies living structure does not just stand alone; it contributes to the life of its surrounding environment, whether that be a neighborhood, a city, or a landscape. This holistic perspective encourages us to consider how our designs and plans will interact with and enhance the natural and social fabric around them. It also requires a sensitivity to the cultural and historical context, ensuring that new structures respect and contribute to the continuity of the built environment.

Alexander's philosophy also critiques the mechanistic approach that often dominates architectural and urban education and practice, where design or plan is segmented into functional, aesthetic, and technical aspects. He argues that this fragmented approach fails to capture the essence of what makes a space feel alive. Instead, he advocates for an organic approach, where buildings are seen as living entities, deeply intertwined with the lives of the people who use them and the environments they inhabit. This perspective encourages us to design with a sense of stewardship, recognizing that our design work is not just about creating isolated structures but about contributing to the ongoing story of human life and the natural world.

In essence, Alexander's living structure invites a fundamental rethinking of architecture as a discipline. It challenges the notion that buildings should be designed solely for efficiency or visual impact and instead emphasizes the creation of spaces that resonate with the deeper needs of human beings and the natural world. This holistic, human-centered approach calls for a return to timeless principles of design, such as levels of scale, local symmetries, thick boundaries, and echoes, which have always underpinned great architecture. By embracing these principles, we can create spaces that are not only functional and beautiful but also profoundly life-affirming—spaces that engage the senses, foster well-being, and contribute to a greater sense of harmony in the world.

As we look towards the future of architecture, Alexander's insights remind us that the most enduring and beloved spaces are those that embody the principles of living structure. These are the spaces that people return to time and again, that become woven into the fabric of their lives, and that stand the test of time not because of their boldness or innovation, but because of their deep-rooted connection to human nature and the natural world. In a world increasingly dominated by rapid technological change and fleeting trends, the concept of living structure offers a path back to the timeless values that make architecture a true art form.

By adopting this perspective, we can move beyond the fragmented, utilitarian practices of the past and towards a more integrated, meaningful approach to design. This approach not only addresses the shortcomings of modernist ideologies but also lays the foundation for a future where the built environment is seen as a living, breathing entity—one that affirms and enriches our daily lives. Through the lens of living structure, architecture can once again become a discipline that serves the deeper needs of humanity, creating spaces that are not only practical and beautiful but also profoundly resonant with the essence of life itself.

## 5. Critique of Modernist Architecture and the Role of Starchitects

Christopher Alexander's critique of modernist architecture targets not just the aesthetic and functional flaws but also the cultural and philosophical forces behind its adoption. A key focus is on "starchitects", architects celebrated for bold and visually striking designs. While influential, they often exacerbate the issues they aim to solve, contributing to disjointed urban environments. Their iconic buildings, though attention-grabbing, can disrupt the urban fabric, creating uninviting spaces disconnected from everyday life. Alexander's critique underscores the need for urban design that prioritizes human experience over architectural spectacle.

### 5.1 The Rise of Starchitects and the Pursuit of Novelty

The 20th century witnessed the emergence of starchitects who, driven by the desire to leave a personal mark on the architectural landscape, prioritized novelty and individual expression over the more enduring values of beauty, functionality, and harmony. This pursuit of the new and the unique became a hallmark of modernist architecture, leading to the creation of buildings that, while visually striking, often failed to meet the needs of the people who lived and worked in them.

Alexander characterizes this phenomenon as a "mass psychosis" (Alexander 2002–2005), a term that reflects the widespread and almost uncritical acceptance of these avant-garde approaches to design. In this context, architects became more concerned with making bold statements and pushing the boundaries of form, often at the expense of creating spaces that were livable, comfortable, and in tune with human scale and needs. This obsession with novelty led to the proliferation of what Alexander describes as "insane, image-ridden, hollow" architecture – buildings that, though they may appear "impressive", are ultimately disconnected from the realities of everyday life.

### 5.2 The Disconnect Between Architecture and Human Experience

One of Alexander's key critiques is that modernist architecture, as epitomized by the work of many starchitects, often results in a built environment that is at odds with the needs and desires of its users. The emphasis on abstract forms, stark materials, and unconventional designs can lead to spaces that feel alienating, cold, and unwelcoming. Instead of enhancing human life, these buildings can create environments that are uncomfortable, disorienting, and even oppressive.

The fundamental purpose of architecture is to create spaces that support and enrich human life. This involves not only meeting functional requirements but also addressing the psychological and emotional needs of the people who inhabit these spaces. Modernist architecture, in its quest for innovation and self-expression, often neglects these aspects, leading to a widespread dissatisfaction with the built environment.

### 5.3 The Role of Starchitects in Perpetuating Architectural Ugliness
Starchitects, with their significant influence and visibility, have played a crucial role in shaping the architectural trends of the 20th and 21st centuries. However, Alexander argues that their work often contributes to what he calls the "uglification" of our cities. The pursuit of iconic status can lead architects to prioritize form over substance, resulting in buildings that, while visually striking, lack the deeper qualities that make a space truly beautiful and meaningful.

Moreover, the influence of starchitects extends beyond individual buildings to the broader cultural and professional context of architecture. Their work sets trends and establishes norms that other architects feel pressured to follow, even when these trends may not serve the best interests of the people who use the buildings. This has led to a homogenization of architectural styles, where the same kinds of stark, abstract forms are replicated across different cities and contexts, contributing to a sense of disconnection and placelessness in the built environment.

### 5.4 A Call for a Return to Human-Centered Design
Alexander's critique is not just a condemnation of modernist architecture and starchitects; it is also a call to action. He advocates for a return to architecture that is grounded in human experience, one that prioritizes the creation of spaces that are not only functional but also deeply resonant with the people who use them. This involves moving away from the pursuit of novelty for its own sake and instead focusing on the timeless principles of beauty, harmony, and wholeness.

In Alexander's view, architects should see themselves not as artists striving for personal expression but as creators of environments that support and enhance human life. This requires a profound shift in how architecture is conceived and practiced, moving away from the abstract, mechanistic approaches of modernism and towards a more organic, human-centered way of thinking about space and design.

### 6. Towards a New Vision of Architecture
Christopher Alexander's vision for the future of architecture represents a paradigm shift, moving away from the fragmented, mechanistic views (Descartes 1644, 1983) that have dominated the field since the advent of modernism. He critiques the reductionist approach that has characterized much of 20th-century architecture, where buildings are often treated as mere objects composed of isolated parts. This approach strips architecture of its deeper meaning and fails to create spaces that truly resonate with the human spirit.

Instead, Alexander advocates for an architectural practice rooted in holistic thinking (Whitehead 1929, Bohm 1980). In his view, a building should not be seen as a collection of separate components—walls, windows, roofs—but as an integrated whole where every element is interrelated and contributes to a larger order. This holistic approach recognizes that the function of a building cannot be divorced from its form, and that ornamentation is not merely decorative but integral to the structure's overall harmony and coherence. This emphasis on the inseparability of function and ornament challenges the minimalist tendencies of modernism, where the pursuit of "form follows function" often leads to stark, lifeless spaces.
Central to Alexander's vision is the concept of living structure, a term he uses to describe spaces that possess a certain quality of life. This idea is not limited to organic forms but applies to all matter, whether natural or man-made. A living structure is one that embodies a sense of order, harmony, and wholeness, making it resonate deeply with those who experience it. The degree to which a building or space is "alive" is determined by its structure—specifically, how its various parts relate to each other and to the whole. A space that is rich in connections, patterns, and coherence will naturally evoke a sense of well-being and comfort in its occupants.

To achieve this vision, Alexander calls for architects to develop a deep understanding of the principles that govern living structures. This involves studying the patterns and relationships (Alexander et al. 1977) found in nature, which have evolved over millennia to create environments that support life. By applying these principles to architectural design documented in his magnum opus *The Nature of Order* (Alexander 2002–2005), we can create spaces that are not only functional but also deeply attuned to human needs—both physical and emotional. This approach requires a shift in focus from designing buildings as isolated objects to viewing them as part of a larger, interconnected whole that includes the environment, the community, and the individual.

Moreover, Alexander's vision emphasizes the importance of creating spaces that resonate with human feelings and experiences. He argues that architecture should go beyond mere utility and aesthetics to engage with the deeper aspects of human existence. A well-designed building, in his view, should evoke feelings of peace, joy, and a sense of belonging (Alexander 1979). It should be a place where people feel connected to themselves, to others, and to the world around them. This requires us to consider not only the physical dimensions of space but also its psychological and emotional impact.

In practice, this means designing spaces that are adaptable, human-centric, and responsive to the changing needs of their users. It involves incorporating elements that foster a sense of continuity with the past while accommodating the realities of the present and future. Alexander's vision thus calls for a new kind of architecture—one that is deeply rooted in the realities of life, sensitive to the subtleties of human experience, and committed to creating environments that support and enhance the well-being of all who inhabit them.

## 7. Conclusion

Alexander's critique of modern architecture is not only a call for reform within the architectural discipline but also a blueprint for rethinking urban management. His concept of living structure, when applied to urban design, offers a path toward cities that are not only functional but also deeply resonant with the human experience. Urban managers and planners can draw on these principles to create sustainable, adaptable cities that nurture both their inhabitants and the environment. Modernist architecture, with its emphasis on cold, machine-like efficiency, standardized forms, and a top-down approach, often neglects the fundamental human need for connection, warmth, and harmony. It is driven by a pursuit of innovation for its own sake, resulting in spaces that are abstract, minimalist, and detached from the lives of those who inhabit them. This approach, while technically advanced, frequently sacrifices the organic richness and contextual sensitivity that are essential for creating environments where people truly feel at home.

In stark contrast, Alexander's vision of architecture is one that is deeply human-centered, rooted in the creation of living structures that resonate with the natural and social rhythms of life. Alexandrian architecture does not prioritize form over function, nor does it treat aesthetics as a mere afterthought. Instead, it seeks a harmonious balance where form and function are inseparable, and where beauty emerges naturally from the context and purpose of the design. This approach emphasizes the maintenance of harmony across scales, from the smallest detail to the overall form, ensuring that every aspect of the built environment contributes to a sense of belonging and vitality.

Alexandrian architecture reclaims the role of architecture as a means of enhancing the human experience. It challenges us to move beyond the sterile, mythological narratives of modernism and to embrace a more organic, knowledge-based approach that is integrated with nature and sensitive to the complexities of human life. Alexander's insights offer a transformative vision for the future of architecture—one that is not only more attuned to the needs of people but also more in harmony with the world around us. By embracing these principles, we can create spaces that are not only functional and aesthetically pleasing but also deeply enriching and life-affirming, setting the stage for a built environment that truly reflects the essence of what it means to be human.

**Acknowledgement:**
This paper was prepared with the support of ChatGPT-4; however, the author takes full responsibility for any errors or oversights. The author gratefully acknowledges the Christopher Alexander & Center for Environmental Structure Archive for providing the images in Figures 1 and 2, with special thanks to Maggie Alexander. Appreciation is also extended to Qianxiang Yao and Huan Qian for their help in creating Figures 3 and 4. Special thanks to Professor Chaosu Li for his insightful comments, which greatly contributed to improving the paper.